\documentclass[journal=jacsat,manuscript=article]{achemso}
\DeclareUnicodeCharacter{202F}{\,}
\usepackage[version=3]{mhchem} 
\usepackage[hidelinks]{hyperref}
\usepackage{textcomp}
\usepackage{color}
\graphicspath{{./Images/}}
\usepackage{mathtools}
\usepackage{graphicx}
\usepackage[dvipsnames]{xcolor}
\usepackage{ulem}
\usepackage{pdfpages}

\makeatletter
\AtBeginDocument{\let\LS@rot\@undefined}
\makeatother

\usepackage{pgffor}
\newcommand{\curlyB}{\mathcal{B}}
\newcommand{\sigmaeff}{\sigma_{\rm eff}}

\author{Hadi Arjmandi-Tash}
\email{hadi.arjmandi@nannovative.com}
\affiliation{Department of Precision and Microsystems Engineering, Delft University of Technology, Mekelweg 2, 2628 CD Delft, The Netherlands}
\alsoaffiliation{Nannovative, Alexander Fleminglaan 1, 2613 AX Delft, The Netherlands}

\author{Roshan Prasad}
\affiliation{Department of Precision and Microsystems Engineering, Delft University of Technology, Mekelweg 2, 2628 CD Delft, The Netherlands}

\author{Hanqing Liu}
\affiliation{Department of Precision and Microsystems Engineering, Delft University of Technology, Mekelweg 2, 2628 CD Delft, The Netherlands}

\author{Gerard Verbiest}
\affiliation{Department of Precision and Microsystems Engineering, Delft University of Technology, Mekelweg 2, 2628 CD Delft, The Netherlands}

\author{Dominic Vella}
\affiliation{Mathematical Institute, University of Oxford, Woodstock Road, Oxford OX2 6GG, United Kingdom}

\author{F. Alijani}
\email{f.alijani@tudelft.nl}
\affiliation{Department of Precision and Microsystems Engineering, Delft University of Technology, Mekelweg 2, 2628 CD Delft, The Netherlands}

\title[Short Title for Header]{
  {\fontsize{16pt}{20pt}\selectfont Mechanical Reinforcement of Graphene via Wrinkling}\vspace{20pt}
}


\begin{document}






\begin{abstract}

Mechanical cantilevers are central to nanotechnology, with ultimate sensitivity achieved at the atomic limit, where low bending rigidity makes stability the fundamental challenge. Here, we introduce a wrinkle-induced stiffening approach that enhances the bending rigidity of monolayer graphene by several orders of magnitude, enabling the fabrication of mechanically robust graphene cantilevers. When suspended over microcavities, these wrinkled membranes exhibit significant increases in both in-plane and out-of-plane stiffness, as confirmed by nanoindentation and resonance measurements, which also reveal that enhanced bending rigidity strongly influences their vibrational response. This behavior marks a transition from tension-dominated mechanics to a regime where bending effects become prominent, even in a single atomic layer. By sculpting these structures, we realize graphene cantilevers with measured bending rigidities between $10^6$–$10^7$ eV, while maintaining femtogram-scale mass. These findings open new directions in nanomechanical sensing and cantilever-based technologies.

\end{abstract}
\section{Introduction}

Although graphene is widely recognized as a two-dimensional (2D) material, its lattice is never perfectly flat, practically. Instead, graphene structure exhibits a spectrum of out-of-plane deformations of distinct characteristics\,\cite{Deng2016}. Thermal fluctuations with subnanometric amplitudes are intrinsic deformations that spontaneously arise within the graphene lattice at finite temperatures\,\cite{Fasolino2007, Los2016}. These dynamic fluctuations are fundamental to the energetic stabilization of graphene, serving as a natural mechanism that maintains its structural integrity. Moreover, they give rise to a number of exotic properties, including negative coefficient of thermal expansion\,\cite{Arjmandi-Tash2017b} and size-dependent elastic moduli\,\cite{Sajadi2018}.

Beyond inherent fluctuations, graphene may exhibit non-intrinsic deformations such as wrinkles, induced by external forces\cite{Nicholl2015a,Ares2021}. Wrinkles are observable in a wide range of amplitudes, from a few to hundreds of nanometers. Smaller--amplitude wrinkles within this spectrum are responsive to external loading. Especially when stretched, such wrinkles would gradually be suppressed (``ironed-out"), a phenomenon that is accountable for the softening effect that is often observed in the stress--strain response of graphene membranes\,\cite{Verbiest2016, Sarafraz2021, Nicholl2015a, Nicholl2017}.

In compressive loadings, on the other hand, small--amplitude wrinkles tend to evolve to generate deformations of higher amplitudes\,\cite{Zang2013, Zhang2014b}. Large--amplitude wrinkles, however, may feature self--contact regions\,\cite{davidovitch2021rucks} with strong van der Waals attractions between the contact surfaces. Such wrinkles are difficult to fully flatten, even after removing the external compression\,\cite{Zang2013}. Practically, these large--amplitude, stable deformations can emerge from a variety of mechanisms, including compression--induced buckling on pre--stretched elastic substrates \cite{Zang2013}, thermomechanical stresses encountered during graphene growth \cite{Ahmad2011}, or mechanical perturbations introduced during handling and transfer processes.

From the technical point of view, the inclusion of wrinkles in graphene structure is expected to improve the mechanical stability of the material. Such irreversible, large--amplitude wrinkles can enhance the out--of--plane rigidity of planar sheets,\cite{Blees2015, Zhou2025a} enabling the formation of stable corrugated structures. In fact, free-standing monolayer graphene structures, hitherto, have been realized only in the form of the circular (perimetral clamped), rectangular (four-side clamped)\,\cite{Zande2010, Castellanos-gomez2015} or doubly-clamped\,\cite{Arjmandi-Tash2017b} membranes. Monolayer graphene cantilevers --- strips fixed at one end and free at the other --- potentially represent the ultimate limit of mechanical softness and minimal mass, making them highly attractive for nanomechanical sensing applications. However, their extreme mechanical fragility renders them inherently unstable, and have thus far eluded experimental realization, except when supported in a liquid environment\,\cite{Blees2015}.

In this work, we overcome this challenge and realize monolayer graphene cantilevers by harnessing wrinkle--induced mechanical reinforcement. We deliberately introduce a network of large--amplitude wrinkles into graphene to boost its bending rigidity and prevent collapse. Our fabrication approach involves a surfactant--mediated lateral contraction of centimeter--scale monolayer graphene at an air--liquid interface. This process generates an interconnected wrinkling pattern across the graphene sheet. The wrinkled graphene is then transferred onto a microstructured substrate containing an array of microcavities, yielding suspended graphene membranes that are clamped on all four sides. These four--edge--clamped membranes serve as a convenient platform to study the influence of wrinkles on mechanical properties. Using nanoindentation, we observe that the presence of wrinkles significantly enhances the membrane’s 2D elastic modulus compared to flat graphene. Interestingly, we also observe an increase in the linear stiffness despite the anticipated softening from lateral contraction. This suggests that the induced out-of-plane deformations have enhanced the importance of bending stiffness compared to the membrane tension in wrinkled samples. 
To confirm this interpretation, we perform resonant measurements on  wrinkled graphene drums. The observed vibrational response reveals a clear departure from tension-dominated dynamics, demonstrating that bending rigidity plays an important role in shaping the resonance characteristics of these membranes. Motivated by these observations, we convert the all-edge clamped membranes into one--end--anchored cantilevers while preserving the wrinkle--induced stiffening. Remarkably, despite being only one atom thick, these wrinkled graphene cantilevers exhibit effective bending rigidities on the order of $10^6-10^7$\,eV --- many orders of magnitude higher than the nominal value of $\kappa \approx 1.2$\,eV, known for pristine graphene \cite{Sajadi2018} --- and even above the thermally renormalized bending stiffness of micron scale graphene. Our cantilevers exhibit ultralow spring constant in the range of $\sim10^{-2}$ N/m. At the same time, the polymer--free fabrication process, allows the graphene to retain its exceptionally low mass (estimated as $\sim 10^{-13}-10^{-14}$\,g), characteristics that place them in an unprecedented regime of mechanical devices. Our findings demonstrate a viable route towards stabilizing atomically thin cantilevers by engineered corrugations. The wrinkle--assisted strategy, proposed here yields ultra--thin, ultra--light, yet robust cantilevers, opening the door to graphene--based mechanical sensors and resonators with unprecedented sensitivity and performance.

\section{Fabrication of Wrinkled Graphene}
\begin{figure}[H]
\centering
    \includegraphics[width=0.8\textwidth]{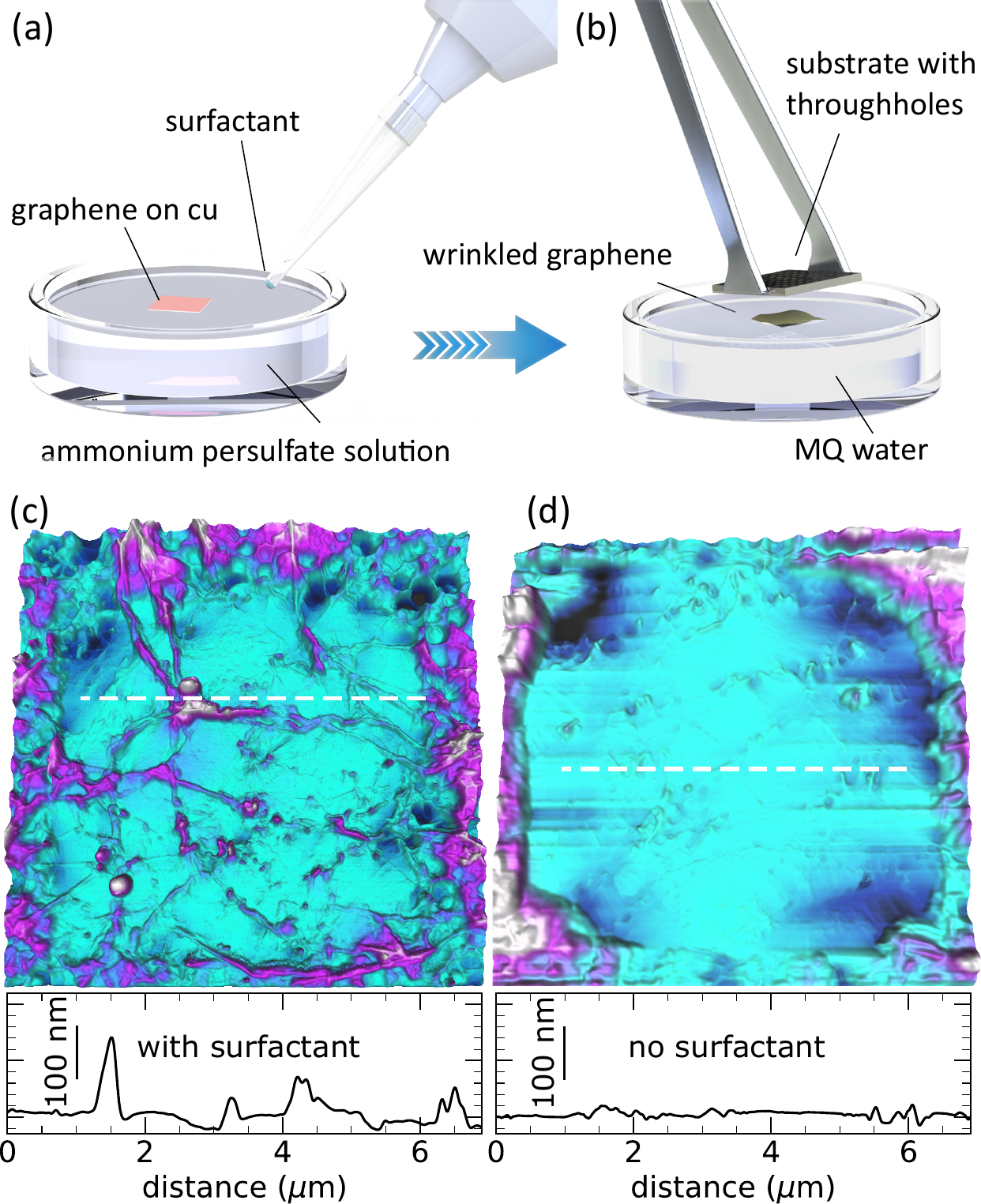}
    \caption{Fabrication and characterization of wrinkled graphene membranes.
a) Graphene grown on copper is floated on a 0.5\,M ammonium persulfate (APS) solution with a trace amount of surfactant (see Supplementary S1). 
b) As the copper etches, surfactant-induced lateral forces generate wrinkles in the graphene. The APS is gradually replaced with pure water, and the wrinkled graphene is transferred onto a substrate with square through-holes and air-dried. 
c) AFM image and profile of a wrinkled (“surfactant-mediated”) graphene membrane suspended over an $8 \times 8\,\mu\mathrm{m}^2$ hole. 
d) Same measurement for a “surfactant-avoided” membrane prepared without surfactant. The color code in (c) and (d) indicates the local out-of-plane deformation of the membrane, with cyan corresponding to regions close to the flat reference plane and purple marking areas with th largest height deviations.}
    \label{fig:fabrication}
\end{figure}

Fabrication of the samples follows a protocol that we developed earlier\,\cite{Lima2018, Arjmandi-Tash2018e}. Briefly, a centimeter-scale graphene sample, chemically grown on a copper foil, was subjected to a lateral force mediated by an enclosing surfactant layer on a liquid surface, after etching away the copper foil. The stress difference at the air-liquid interface contracts the surface area of graphene to form a random network of wrinkles/folds, associated with a detectable reduction in the graphene area. We transferred the wrinkled graphene onto a substrate with a dense array of square through-holes (side length $8\,\mu\text{m}$, corresponding to an effective circular radius $R_\text{out} = \tfrac{8}{\sqrt{\pi}}\,\mu\text{m}$) by placing the substrate in contact with the graphene. \autoref{fig:fabrication}-a and b schematically illustrate the fabrication process, with the details provided in the supporting information S1. The process achieves an array of freestanding membranes which we will refer to as the ``surfactant-mediated" membranes in this text. For the sake of comparison, we fabricated a separate set of the membranes,  referred to as ``surfactant-avoided" membranes,  with no surfactant involved in the fabrication process. 

\subsection{AFM Nanoindentation}
To characterize the topography and probe the mechanical properties of graphene samples, we performed a series of scanning and nanoindentation experiments using an AFM (JPK Nanowizard 4) with an indenter radius of $R_{\text{in}} \approx 10\,\text{nm}$, detailed in the supporting information S2. \autoref{fig:fabrication}-c and d compare the topography of the arbitrarily selected ``surfactant-mediated"  and ``surfactant-avoided" membranes. The presence of a network of interwoven wrinkles/folds is the salient feature of the ``surfactant-mediated" membrane. The amplitude of such wrinkles can easily exceed several tens of nanometers. The ``surfactant-avoided" membrane, on the other hand, still showcases a level of wrinkles, albeit very dilute and with lower amplitudes. Such wrinkles could have been generated for example, due to the liquid sloshing during the rinsing and/or top-fishing of the graphene. Regardless of the amplitude and density of the wrinkles, the AFM mappings demonstrate the propensity of graphene for stable foldings.

\begin{figure}[H]
\centering
\includegraphics[width=.9\textwidth]{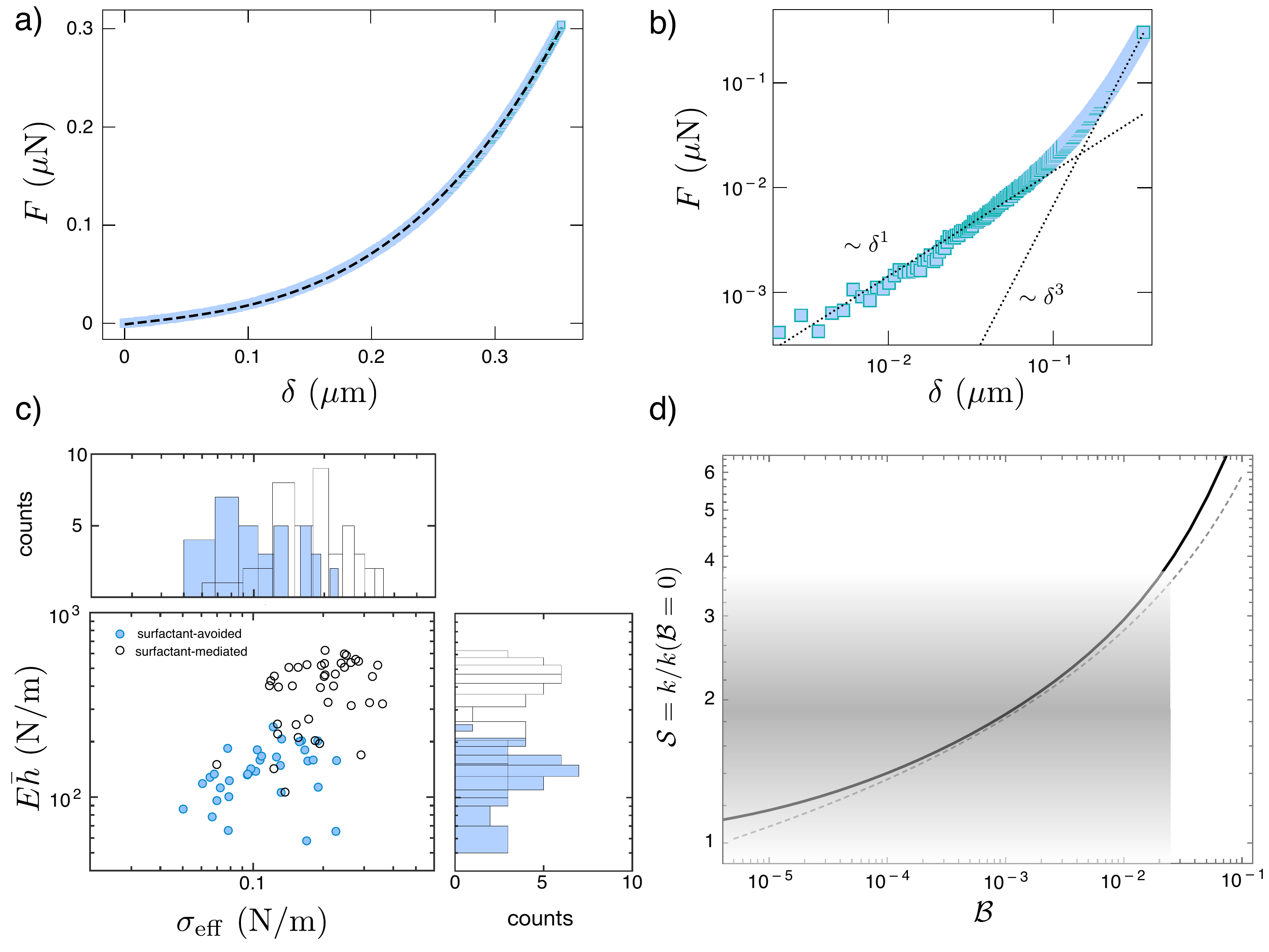}
\caption{Nanoindentation of wrinkled graphene membranes. a) Representative $F-\delta$ correlation, measured for one of the wrinkled-graphene membranes. The broken line is the best fit of the experimental data with \autoref{eq:eq1}. b) The same data plotted on doubly-logarithmic scales shows  the linear and cubic relationships at small and large indentations, respectively. c) Correlation between the estimated $\sigmaeff$ and the $E\bar{h}$ values: histograms of the pretension and the $E\bar{h}$ values (in log scale) are plotted in the side panels. d) Enhancement of the indentation stiffness by the presence of bending stiffness when $R_{\rm in}/R_{\rm out}=2.2\times10^{-3}$. The analytical result for the indentation stiffness, valid for all $\curlyB$ (see supporrting information S3), is rescaled by the value when $\curlyB=0$ to show the  stiffness enhancement, $\mathcal{S}$, that comes from  $\curlyB>0$ (solid curve). When $R_{\rm in}^2/R_{\rm out}^2\ll\curlyB\ll1$, $k$ is given approximately by Equation \eqref{eqn:kIndentMainText} (used to generate the dashed curve).  The shaded region represents the range of indentation stiffening observed in our experiments, with darker regions being more likely.}\label{fig:f_d}
\end{figure}

Nanoindentation provides a quantitative measure of the elastic properties of graphene membranes. Briefly, the membranes are pushed in the center with an AFM tip and the corresponding indentation depth is deduced from the AFM piezoelectric base response, albeit after post-processing and correction for the deflection of the AFM cantilever. Figure~\ref{fig:f_d}a illustrates the correlation between the applied force ($F$) and the corresponding indentation (displacement, $\delta$) at the center of a selected membrane. We note that the work done by the applied force is stored in the bending and stretching energies of the graphene sheet and lead to the following scaling relationship \cite{chandler2020indentation, Vella2017}:
\begin{equation}
F \cdot \delta \sim 
R_{\text{out}}^2 \sigma_{\text{pre}} \left( \frac{\delta}{R_{\text{out}}} \right)^2
+ 
R_{\text{out}}^2 E\bar{h} \left( \frac{\delta}{R_{\text{out}}} \right)^4
+ 
R_{\text{out}}^2 \kappa \left( \frac{\delta}{R_{\text{out}}^2} \right)^2.
\label{eq:eq1}
\end{equation}
where $\sigma_{\text{pre}}$, $E\bar{h}$, and $\kappa$ are the pretension, 2D Young's modulus, and bending rigidity of the graphene sheet, respectively.  Equation \eqref{eq:eq1} involves two terms that are quadratic in the indentation depth $\delta$: one coming from the pre-tension of the sheet, the other from the sheet's bending stiffness. To compare the relative importance of these two effects, we introduce a dimensionless bending stiffness, defined as $\curlyB = \frac{\kappa}{\sigma_{\rm pre} R_{\rm out}^2}$.

Equation~\eqref{eq:eq1} further reveals two distinct regimes in the force–displacement response of the membrane. In the regime of small applied forces, the indentation scales linearly with force, i.e., $F \sim \delta$. In this limit, the membrane’s resistance to the applied normal load is primarily governed by its in-plane pretension and/or its bending rigidity. A detailed analysis \cite{chandler2020indentation} shows that which of these dominates depends on the parameter $\curlyB$: when ${\curlyB}\ll R_{\text{in}}^2/R_{\text{out}}^2$, the bending contribution is entirely negligible, and Equation~\ref{eq:eq1} simplifies to a tension-dominated model. For pristine graphene membranes, with a typical bending rigidity of $\kappa \approx 1.2$ eV, this condition is generally satisfied, and the linear part of the force–displacement curve is commonly used to extract the in-plane pretension $\sigma_{\text{pre}}$\cite{Lee2008,Sajadi2017a, Davidovikj2017, Siskins2025}. However, it has also been suggested\cite{Kosmrlj2013} that wrinkling or thermal crumpling of graphene may lead to an enhanced bending modulus. In this case, detailed analysis shows that the bending stiffness first plays a role when $\curlyB \approx R_{\text{in}}^2/R_{\text{out}}^2$ --- while the indentation stiffness is still proportional to the tension in this limit, the relevant prefactor has a logarithmic dependence on the bending stiffness parameter, $\curlyB$ \cite{chandler2020indentation}. When $\curlyB\gg R^2_{\rm in}/R^2_{\rm out}$ the pretension is negligible and it is the bending stiffness that dominates; this regime has not been reported in monolayer graphene experimentally. 

 To account for the combination of bending stiffness and pretension effects, we define an effective  pretension  parameter $\sigmaeff$, which captures the total linear resistance of the sheet to indentation that is measured experimentally and incorporates contributions from both the actual pretension $\sigma_{\text{pre}}$ and the bending rigidity $\kappa$. In the second, so-called “cubic regime,” observed at high applied forces, the sample stiffens. Here,  larger force increments are required to produce the same indentation increment, obeying: $F\sim\delta^3$. The applied force in the cubic regime stretches the chemical bonds between the carbon atoms in the basal plane; hence, the Young's modulus of the material is the dominant source of the nonlinear stiffness. These two regimes can be well-identified in  \autoref{fig:f_d}a,b for a wrinkled membrane.  In \autoref{fig:f_d}a, the black dashed curve corresponds to the best fit of a representative experimentally measured force-displacement curve with the model in \autoref{eq:eq1}, incorporating dimensionless prefactors appropriately.\cite{Vella2017} The values of the effective tension  $\sigmaeff$, and nonlinear stretching stiffness, $E\bar{h}$,  can be extracted by fitting the experimental nano-indentation curves. In total, 34 membranes of ``surfactant-mediated" and 40 membranes of ``surfactant-avoided" types have been measured. 

Histograms of the estimated $\sigmaeff$ and $E\bar{h}$ are plotted in \autoref{fig:f_d}c. Accordingly, we achieved an average $\sigmaeff$ of $0.12 \pm 0.05\mathrm{~N/m}$ for the ``surfactant-avoided'' sample, which increased to $0.20 \pm 0.07\mathrm{~N/m}$ upon the inclusion of wrinkles: the membrane is apparently stiffened by a factor $\mathcal{S}\approx1.7$ (or $0.8\leq\mathcal{S}\leq 3.9$ accounting for uncertainties). The nonlinear stiffness is also prone to wrinkle-induced changes, increasing from $E\bar{h} = 140 \pm 55\,\text{N/m}$ in the ``surfactant-avoided'' sample to $E\bar{h} = 422 \pm 160\,\text{N/m}$ in the ``surfactant-mediated'' samples. 

While changes in  $E\bar{h}$ can be explained via nonlinear membrane stretching, a closer look at the linear regime reveals a more surprising effect: possible emergence of bending rigidity as a contributor to the nanoindentation response. More specifically, the observed increase in linear stiffness is counterintuitive: the inclusion of surfactant induces lateral contraction of the membrane \cite{kumar2020stresses}, which might be expected to reduce the effective tension, and hence reduce the stiffness. Instead, we observe an enhancement, pointing to a  contribution from wrinkle-induced out-of-plane reinforcement --- an increase in the effective bending stiffness. A possible alternative hypothesis is that the adhesion between the folded regions leads to localized stretching of suspended regions between wrinkles, thereby enhancing the apparent stiffness. However, detailed theoretical analyses of  such out-of-plane deformations show that these large-amplitude wrinkles primarily relieve in-plane stresses, with the response governed by bending rather than by tension buildup \cite{davidovitch2021rucks}. We therefore consider it unlikely that localized stretching is a mechanism behind the observed stiffening.

To quantify the increase in indentation stiffness caused by a non-negligible bending rigidity, we first note the simpler result previously calculated for  a point indenter\cite{chandler2020indentation} for which the indentation stiffness is 
\begin{equation}
\frac{k}{2\pi\sigma_{\rm pre}}\approx\left(\gamma-\log2-\frac{1}{2}\log\curlyB\right)^{-1}
\label{eqn:kIndentMainText}
\end{equation} when $\curlyB\ll1$. Here $\gamma\approx 0.577$ is the Euler–Mascheroni constant. Crucially, the indentation stiffness $k$ depends linearly on the true pretension in the system, $\sigma_{\rm pre}$, and is only logarithmically dependent on $\curlyB$ --- a dependence that is shown in \autoref{fig:f_d}d and that persists in more detailed calculations (see supporting information S3.1). Nevertheless, from the observed stiffening $\mathcal{S}$ of the surfactant-mediated membranes (compared to the surfactant-avoided ones) we can obtain a first estimate of the size of $\curlyB$ needed to give such a large stiffening; we find that 
\begin{equation}
    \curlyB\approx \frac{e^{2\gamma}}{4}\left(\frac{R_{\rm in}}{R_{\rm out}}\right)^{2/\mathcal{S}}\approx5.9\times 10^{-4},
    \label{eqn:CurlyBEstimate}
\end{equation} which represents an enhancement of the bending stiffness by a factor of around $10^4$ compared to the pristine value, but with large uncertainty because of the wide range of values of $\mathcal{S}$. Note, in particular, that the exponential dependence on $\mathcal{S}$ in \eqref{eqn:CurlyBEstimate} means that even a moderate relative stiffening, $\mathcal{S}$, requires an enormous increase in the bending stiffness $\curlyB$, and hence $\kappa$. Considering that the pre-tension is expected to become lower \cite{kumar2020stresses} or at least remain unchanged between the ``surfactant-avoided" and ``surfactant-mediated" cases, we select a representative value \cite{sarafraz2024quantifying} of \( \sigma_{\text{pre}} \sim 0.1 \, \text{N/m} \)  to estimate $\kappa$. Based on this, we estimate that the bending rigidity must lie in the range $8\times 10^3\mathrm{~eV}\leq \kappa\leq 4\times 10^5\mathrm{~eV}$ to be consistent with the observed force–displacement response. A more detailed estimate based on taking full account of the bending stiffness and finite indenter radius gives  $\kappa= 2\times 10^5\mathrm{~eV}$ and is provided in the supporting information S3.1. This value is several orders of magnitude greater than the bending rigidity of pristine graphene.

Further insight can be gained by comparing the nonlinear stiffness $E$\(\bar{h} \) between the two cases. For the ``surfactant-avoided'' samples, the measured values fall below the generally accepted 2D Young’s modulus of pristine graphene ($E$\(\bar{h} = 340 \text{~N/m} \))\,\cite{Lee2008}. This difference aligns with prior observations\,\cite{Nicholl2015a, Sarafraz2021} and is attributed to static ripples, low-amplitude out-of-plane undulations, that are flattened during indentation, costing negligible  stretching energy. In contrast, the wrinkles in the ``surfactant-mediated'' samples are of much larger amplitude, as confirmed by AFM images in \autoref{fig:fabrication}-c,d. These large wrinkles often form self-contacted regions that are mechanically robust and energetically expensive to deform due to van der Waals adhesion\,\cite{davidovitch2021rucks}.

\subsection{Wrinkled Graphene Resonator}

\begin{figure}[H]
\centering
\includegraphics[width=1\textwidth]{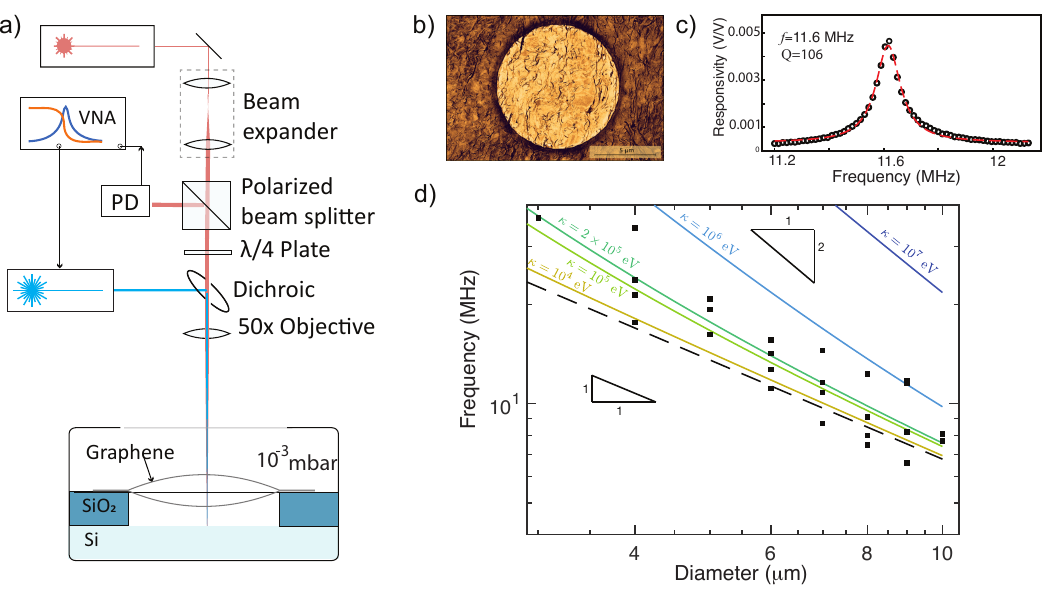}

\caption{Resonance frequency measurement of wrinkled graphene drums. a) Schematic of the interferometric measurement setup used to record the vibrations. PD and VNA stand for Photodiode and Vector Network Analyzer, respectively. b) SEM image of a wrinkled graphene drum. c) Resonant response of a wrinkled drum with an optothermal drive of 0 dbm. The figure also incudes a Lorentzian fit to the data shown in red. d) Diameter-dependent resonance frequencies together with fits from the theoretical model for different values of bending rigidity $\kappa$. Here, results are shown with $\rho \bar{h}=10^{-5}\mathrm{~kg/m^2}$ and $\sigma_{\rm pre}=0.1\mathrm{~N/m}$; similar plots with different values of these parameters are shown in supporting information Fig.~S5.  
}  
\label{fig:resonance} 
\end{figure}

To further investigate the presence of bending rigidity in wrinkled graphene membranes, we measured the fundamental resonance frequency of ``surfactant-mediated'' samples suspended over circular cavities (3--10\,$\mu$m in diameter, 285\,nm deep) etched into a SiO$_2$ substrate. The measurements were performed in vacuum (10$^{-3}$\,mbar) at room temperature using a Fabry--Pérot interferometric setup where a power-modulated blue laser ($\lambda = 405$\,nm) thermomechanically actuated the membranes, while a red laser ($\lambda = 633$\,nm) detected their vibrations. The reflected intensity was then recorded with a photodiode and analyzed via a vector network analyzer (VNA) synchronized with the actuation signal (see \autoref{fig:resonance}a) \cite{kecskekler2021tuning}. A representative SEM image of a wrinkled graphene drum is shown in Figure~\ref{fig:resonance}b.

Previous studies have shown that pristine graphene drums, of diameter $d_\text{eff}=2 R_{\text{out}}$, behave as tension-dominated membranes, with fundamental resonance frequency \( f = \frac{2.4}{\pi d_\text{eff}}\sqrt{ \frac{\sigma_{\text{pre}}}{\rho \bar{h}} } \) where $\rho\bar{h}$ is the mass per unit area of the drum\cite{davidovikj2016visualizing,sarafraz2024quantifying}. In contrast, bending-dominated plates follow \( f = \frac{20.42}{\pi d_\text{eff}^2} \sqrt{ \frac{\kappa}{\rho \bar{h}} } \) \cite{castellanos2013single}. The different scalings of resonance frequency with diameter makes resonance measurements a sensitive probe for identifying the dominant restoring forces --- especially in light of the nanoindentation results discussed earlier which suggested the contribution of bending rigidity to the linear stiffness should not be neglected. To that end, we performed frequency sweep across drums with a variety of diameters and fitted the resulting spectra using a harmonic oscillator model \cite{wopereis2024tuning}. As a representative example, a 7\,$\mu$m-diameter drum showed a resonance frequency of 11.6\,MHz with a Q-factor of 106 (see Figure ~\ref{fig:resonance}c). The Q-factor  value is comparable to the literature reports for suspended graphene \cite{steeneken2021dynamics} and is indicative of moderate damping under vacuum conditions; Q-factor variation across diameters is detailed in the Supplementary  Information Fig S3. The broader dataset (Figure~\ref{fig:resonance}d) shows resonance frequencies ranging from 6.7\,MHz ($d_\text{eff}$ = 9\,$\mu$m) to 37\,MHz ($d_\text{eff}$ = 3\,$\mu$m), revealing a scaling trend that lies between the ideal membrane and plate limits (see \autoref{fig:resonance}d).

To interpret this intermediate behavior, we developed a theoretical model that accounts for finite bending rigidity in membrane oscillations (see supporting  information S3.2) and used it to fit experimental data. Assuming a pre-tension of $\sigma_\text{pre}= $0.1\,N/m---independently extracted from AFM nanoindentation for `surfactant-mediated'' samples and assuming that pre-tension remains unaltered when surfactant is added to the liquid--air interface, we find that the measured frequencies are best explained by bending rigidity values in the range of $10^{5}$--$10^{6}$\,eV. 
The optimal fit corresponds to $\kappa \approx 2 \times 10^{5}$\,eV for a mass density of $ \rho\bar{h}=10^{-5}$ kg/m$^2$ significantly exceeding the theoretical value for pristine graphene, and closely matching the value obtained from nanoindentation experiments. These results suggest that the wrinkled graphene drums operate in a hybrid mechanical regime, where both pre-tension and bending rigidity contribute to the observed dynamics.

\subsection{Wrinkled Graphene Cantilever}
\begin{figure}[H]
\thispagestyle{empty}
\centering
\includegraphics[width=0.95\textwidth]{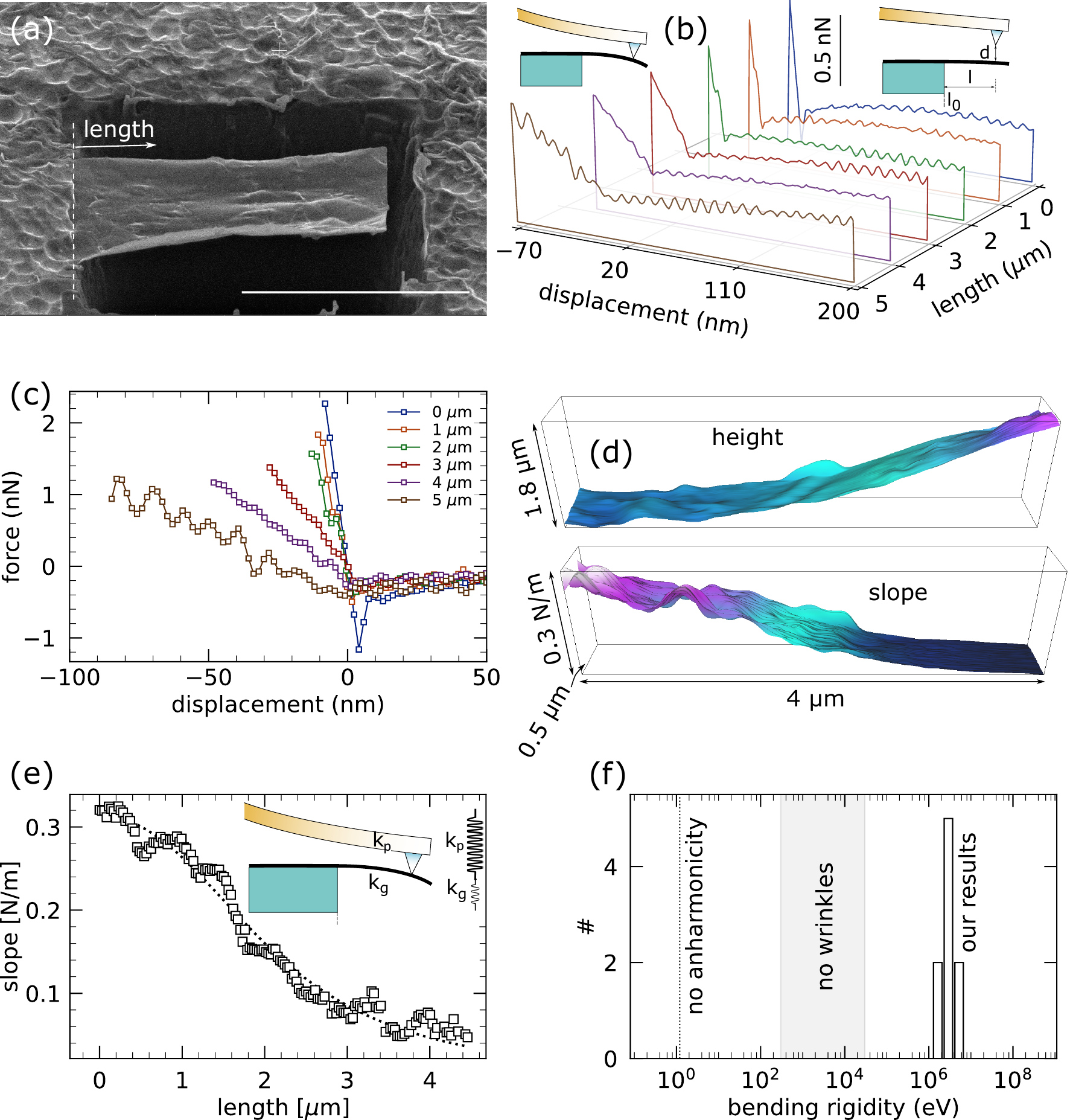}
\caption{Characterization of the wrinkled-graphene cantilevers: a) Scanning electron microscopy of a graphene cantilever; The scale bar corresponds to 5$\,\mu$m. b) Force-displacement relation of the graphene cantelever in (a), obtained by nanoindentation at different lengths. c) Comparison of the slopes of the curves in the contact region in (b); d) 3D height (top) and slope (bottom) profiles measured along the same cantilever. e) Fitting of the measured slope along the graphene cantilever with the model of \autoref{eq3}; The data points are the average over the width of the 0.5$\,\mu$m stripe in the bottom figure in (d). $k_p$ and $k_g$ in the inset schematic refer to the bending rigidities of the probe and the graphene cantilever, respectively. f) Bending rigidities of the wrinkled graphene cantilevers (our results), benchmarked with the published data. ``No anharmonicity" corresponds to the theoretical value of the bending rigidity ($\kappa$ = 1.2\,eV) of graphene samples. ``No wrinkles" corresponds to the microscale graphene samples experimentally measured previously,\cite{Blees2015} with no extra wrinkles included.}
\label{fig:cantilever}
\end{figure}

Along with the increased in-plane stiffness, the inclusion of the wrinkles is expected to make graphene stiffer in the out-of-plane direction as shown by AFM nanoindentation and resonant measurements. However, while these methods show clearly that the effective bending stiffness of the surfactant-mediated graphene is significantly increased compared to, for example, pristine graphene, the uncertainty in the values of other parameters (notably $\sigma_{\rm pre}$ and $\rho\bar{h}$) make determining a precise estimate of $\kappa$ difficult. Within this perspective we sought to make a direct measure of the bending rigidity; we sculpted the wrinkled membranes --- by removing the excess areas with a focused ion beam (see supporting information S1 for more details) --- and realized graphene cantilevers. We fabricated a total of nine cantilevers with lengths in the range of 5 to 7\,$\mu$m and widths in the range of 2 to 3\,$\mu$m. \autoref{fig:cantilever}-a provides the scanning electron micrograph (SEM) of a selected cantilever. We analyzed the bending rigidity of the graphene cantilevers further with an AFM nanoindentation in detail. Unlike fully clamped membranes, where internal tension affects the stiffness, the response of cantilevers is dominated by bending rigidity, making them a more direct probe of this property. \autoref{fig:cantilever}-b illustrates a set of the force-displacement curves, measured over multiple points along the length of a graphene cantilever (note the definition of the ``length" in \autoref{fig:cantilever}-a). The ``approach" and ``contact" regions of the curves can be respectively identified at the positive and negative sides of the ``displacement" axis. \autoref{fig:cantilever}-c provides a closer look into the contact region of the curves. Intuitively, it is expected that the stiffness of the graphene cantilevers decreases by increasing the length; this is translated into less steep curves in the contact region. Unlike the indentation of the fully-clamped membranes (\autoref{fig:f_d}-a), where the force grows by higher orders of the displacement (reaches the third order at sub-micron displacements), the displacements in cantilever structures are in much smaller ranges (tens of nanometers) and preserve linear dependencies with constant, but length-dependent slopes. Importantly, the diminishing slope is evident at larger lengths along the cantilever.

To minimize the effect of local perturbations by the wrinkles in the later analysis, we mapped the graphene cantilevers in  windows of width 0.5\,$\mu$m along the length, using the AFM set-up. The inferred ``height" and ``slope" profiles are plotted in \autoref{fig:cantilever}-d. The process of sculpting induced
an up-curvature in the majority of the cantilevers. Particularly for this sample, the cantilever curled up by $\sim$\,1.8\,$\mu$m at the length of 4\,$\mu$m. The diminishing slope of the ``contact region" --- locally perturbed by the wrinkles --- is evident in the bottom mapping.

Similar to the indentation of the fully clamped membranes, the spring constants of the AFM probe and graphene cantilever fall in a series configuration for the combined system (see the inset schematics in \autoref{fig:cantilever}-e). The overall spring constant reads as:

\begin{equation}
k_{\text{tot}}^{-1}={k_\text{p}}^{-1}+{k_\text{g}}^{-1},
\label{eq3}
\end{equation}

\noindent with $k_\text{p}$ and $k_\text{g}$ referring to the spring constants of the probe and graphene cantilevers, respectively. We measured the constant $k_\text{p}$ by pushing the AFM cantilever against the rigid substrate. The spring constant of the graphene cantilever, however, strongly decays with the length, as $k_\text{g} = {3\kappa w}/{l^3}$, with $\kappa$, $l$, and $w$ as the bending rigidity, the length, and the width of the wrinkled graphene cantilever. Considering the series spring model with length-dependent $k_\text{g}$, the measured overall spring constant of the system ($k_\text{tot}$) approaches $k_\text{tot}\rightarrow k_\text{p}$ and $k_\text{tot}\rightarrow 0$  in the extremes of $l\rightarrow 0$ and $l \gg 0$, respectively. Note that for the analyses in this section, the zero displacement ($d = 0$) is defined as the point along the length where the slope starts to deviate from the measured stiffness of the AFM cantilever ($\sim$ 0.3\,N/m).\autoref{fig:cantilever}-e fits the experimentally measured $k_\text{tot}$ (i.e., the slope of the contact region in \autoref{fig:cantilever}-b and c) with the model of \autoref{eq3}. The presented slope (the vertical axis) in this plot is the average over the widths (0.5\,$\mu$m) of the mapped window, as in \autoref{fig:cantilever}-d. Such fittings provide estimates for the bending rigidity $\kappa$, as summarized in \autoref{fig:cantilever}-f.

Theoretically, a monolayer flat graphene sheet is postulated to show an ultimately small bending rigidity, estimated by neglecting the effect of the intrinsic thermal fluctuations of the lattice\,\cite{Fasolino2007, Los2016}. At finite temperatures, anharmonic coupling between in-plane and out-of-plane thermal fluctuations renormalizes the bending rigidity as 
$\kappa(l) \sim \kappa\, l^{\eta}$, where $l$ is the characteristic length scale, $\kappa$ is the bending rigidity of pristine graphene and $\eta \approx 0.8$~\cite{Nelson1987}. In fact, it has been experimentally shown that the thermal fluctuations in micro-scale samples (with no wrinkles involved) can stiffen the graphene by up to five orders of magnitude\cite{Blees2015} (the grey-shaded region in \autoref{fig:cantilever}-f). Our samples, with deliberately included wrinkles, on the other hand, showcase $\kappa$ in the range of 10$^6$ to 10$^7$\,eV, which is reasonable considering the combined hardening effects of the thermal fluctuations and large--amplitude wrinkles. These findings are consistent with our estimates of bending rigidity from the AFM nanoindentation as well as resonant measurements of all edge clamped square membranes.

Mechanical cantilevers offer ultrasensitive measurements across fields ranging from material science to diagnostics~\cite{li2007ultra,boisen2011cantilever}. Recent advances in miniaturization have led to devices that are only a few nanometers thick, such as Al$_2$O$_3$ cantilevers~\cite{Zhou2025a,Zhou_2025}. Yet, their ultimate performance can only be achieved when the thickness is shrunk to the atomic limit, a regime where maintaining mechanical stability poses a fundamental challenge.  The realization of monolayer graphene cantilevers, hitherto, has met the challenge of the instability of atomically thin  cantilevers due to the diminishing bending rigidity associated with the extremely low thickness of the material. Our approach of increasing the bending rigidity by the inclusion of a finite level of wrinkles stands out in this respect. The graphene cantilevers reported in our work represent unprecedented examples of ultra-light, ultra-thin, and soft, yet stable, mechanical structures. Benchmarking the characteristics of the graphene cantilevers with the known system of AFM cantilevers is interesting here. The spring constant of a typical AC mode AFM cantilever (NCLR Nanoworld is arbitrarily selected here) with a 200\,$\mu$m length is in the range of a few tens of N/m, while a graphene cantilever with only 5\,$\mu$m length exhibits a spring constant of only 10$^{-2}$\,N/m. The bending rigidity of such an AFM cantilever, on the other hand, is on the order of 10$^{13}$\,eV; the value drops by up to seven orders of magnitude for graphene cantilevers, reaching 10$^{6}$\,eV. Interestingly, the mass of the graphene cantilevers—achieved from a mono-atomic starting material—is astonishingly low. With a mass density of \( \rho \bar{h} \approx 10^{-{5}} - 10^{-{6}} \, \text{kg/m}^2 \)\,\cite{steeneken2021dynamics}, we estimate a mass of \( 10^{-{13}} - 10^{-{14}} \, \text{g} \) for the cantilever shown in \autoref{fig:cantilever}, which is orders of magnitude lower than that of typical AFM cantilevers (\( \sim 10^{-7} \, \text{g} \)).  Consequently, free-standing graphene cantilevers can access a previously unexplored mechanical regime, enabling measurements at the very limits of force, mass, and displacement detection.

\section{Conclusion} 

In this study, we characterized the effect of wrinkles on the elastic properties of graphene and highlighted their potential for practical applications. We fabricated different types of samples --- fully clamped drums and cantilevers --- and examined them using a nanoindentation approach as well as resonant measurements. Indentation measurements of complete, but wrinkled, membranes revealed that their mechanical response is controlled by a combination of tension and bending rigidity --- an effect that is only possible with a significantly increased bending rigidity compared to pristine graphene. Complementary resonant measurements further revealed a clear deviation from the tension-dominated behavior, exhibiting resonance characteristics indicative of a hybrid mechanical regime in which both pre-tension and finite bending rigidity play a  role. The combination of these experiments provided a comprehensive picture of the enhanced mechanical performance of wrinkled graphene sheets. Moreover, wrinkling graphene increased its bending rigidity enough to stabilize the structure as a cantilever. This allows the creation of extremely soft springs while maintaining the low mass of a single atomic layer thereby paving the way for the use of graphene cantilevers as ultrasensitive NEMS devices. 

\section{Data availability}
The data that support the findings of this study are available from the corresponding authors upon reasonable request.

\section{Acknowledgment}
This project has received funding from European Union’s Horizon 2020 research and innovation programme under Grant Agreement Nos. 802093 (ERC starting grant ENIGMA) and 101125458 (ERC Consolidator grant NCANTO). Views and opinions expressed are however those of the author(s) only and do not necessarily reflect those of the European Union or the European Research Council. Neither the European Union nor the granting authority can be held responsible for them. Authors acknowledge fruitful discussions with Peter Steeneken, Ali Sarafraz, and Benjamin Davidovitch.

\section{Authors contribution}
{H.A., G.V., D.V., and F.A. conceived the experiments and methods; H.A., and R.P. fabricated the graphene samples; R.P., H.A., and H.L. conducted the measurements and the experimental data was analyzed by H.A., G.V., D.V., and F.A.; D.V. further built the theoretical model; F.A. supervised the project; and the manuscript was written by H.A., D.V., and F.A. with inputs from all authors.}

\section{Competing interests}The authors declare no competing interests.

\bibliography{library_used}


\bigbreak


\clearpage



\section*{Supplementary material}

# Supporting Information: Mechanical Reinforcement of Graphene via Wrinkling

Hadi Arjmandi-Tash[*,1,2], Roshan Prasad[1], Hanqing Liu[1], Gerard Verbiest[1], Dominic Vella[3], and Farbod Alijani[*,1]

[1]Department of Precision and Microsystems Engineering, Delft University of Technology, Mekelweg 2, 2628 CD Delft, The Netherlands
[2]Nannovative, Alexander Fleminglaan 1, 2613 AX Delft, The Netherlands
[3]Mathematical Institute, University of Oxford, Woodstock Rd, Oxford OX2 6GG, UK

August 22, 2025

## S1. Fabrication of Wrinkled Graphene Membranes and Cantilevers

The fabrication of wrinkled graphene samples was achieved through the action of surfactants on a liquid subphase. Briefly, a piece of chemically grown graphene on a copper foil (sourced from Graphenea), approximately $1\times1$ cm$^2$ in size, was back-etched to remove unwanted graphene from the foil's backside and then placed on the surface of a $0.5$ M ammonium persulfate (APS) solution in a petri dish. Next, a small amount of surfactant ($20\,\mu$l of 0.1% Liquinox, commonly used for cleaning glassware in chemical labs) was added to the subphase to lower the surface tension for the "surfactant-mediated" sample. A control sample, referred to as the "surfactant-avoided" sample, was prepared in the same way but without the addition of surfactant.

Over time, the APS solution gradually etched away the copper foil in both samples, leaving the graphene sheets floating at the air-liquid interface. The presence of surfactants in the "surfactant-mediated" sample induced the formation of wrinkles, leading to an estimated 30% reduction in the sample area. Next, the APS subphase was replaced with Milli-Q (MQ) water using a syringe, and the process was repeated several times to remove any remaining APS salts from the graphene. Finally, both samples were transferred onto identical chips with square-shaped through-holes ($8\times8$ $\mu$m$^2$) as well as circular cavities etched in 285 nm silicon oxide with diameters ranging from $3\mu$m to $10\mu$m by gently placing the chips in contact with the floating graphene sheets. The samples transferred on through-holes were particularly used for AFM nanoindentation measurements, while the samples transferred over circular cavities were used to measure the vibrations of suspended wrinkled graphene.

Furthermore, to fabricate graphene cantilevers, the same wrinkled membranes were sculpted using Focused Ion Beam (FIB) on three sides (see Fig S1). Overall nine cantilevers with lengths in the range of 5 to 7 $\mu$m and widths in the range of 2 to 3 $\mu$m. Pictures of some of these cantilevers is brought in Fig S2.

## S2. Characterization of Wrinkled Graphene

Atomic force microscopy (AFM) was the primary characterization technique used to analyze both the topography and mechanical properties of the graphene samples. The experiments were conducted using a JPK NanoWizard 4 XP BioScience instrument from Bruker. The sample surfaces were scanned using AC and/or Tapping® modes to assess topography and pre-select suspended, defect-free membranes and cantilevers. This scanning process provided insights into the geometry of the samples, the degree of wrinkling, and the presence of defects such as holes or tears before proceeding with further mechanical testing.

The mechanical properties of the membranes and cantilevers were investigated using the nanoindentation approach [3] in QI$^{\text{TM}}$-Advanced mode. Fully clamped membranes were indented at their center, while cantilevers were loaded along predefined stripes along their length with a preset indenting force. The corresponding deflections of the samples were recorded during loading, and the actual deflections were extracted by excluding cantilever bending during post-processing. Depending on the scanning mode, we used appropriate AFM probes: NCLR Nanoworld cantilevers ($f_0 \approx 170$ kHz, $k \approx 30$ N/m) for AC scanning mode, and CONT Nanoworld cantilevers ($f_0 \approx 13$ kHz, $k \approx 0.3$ N/m) for Tapping and QI$^{\text{TM}}$-Advanced modes, where $f_0$ and $k$ represent the measured resonance frequency and spring constant of the probes, respectively.

Furthermore, to perform resonant measurements on wrinkled membranes suspended over circular cavities, as explained in the main text, a Febry-Perot interferometery set-up was used and actuation was provided by using a

blue laser diode that was power-modulated via a Vector Network Analyzer (VNA), and thermomechanically drove the membranes in motion (see [4] for details). A total of 28 "surfactant-mediated" samples were measured at 0dBm drive level. The measured resonances as well as their Q-factors ere brought in Fig S3.

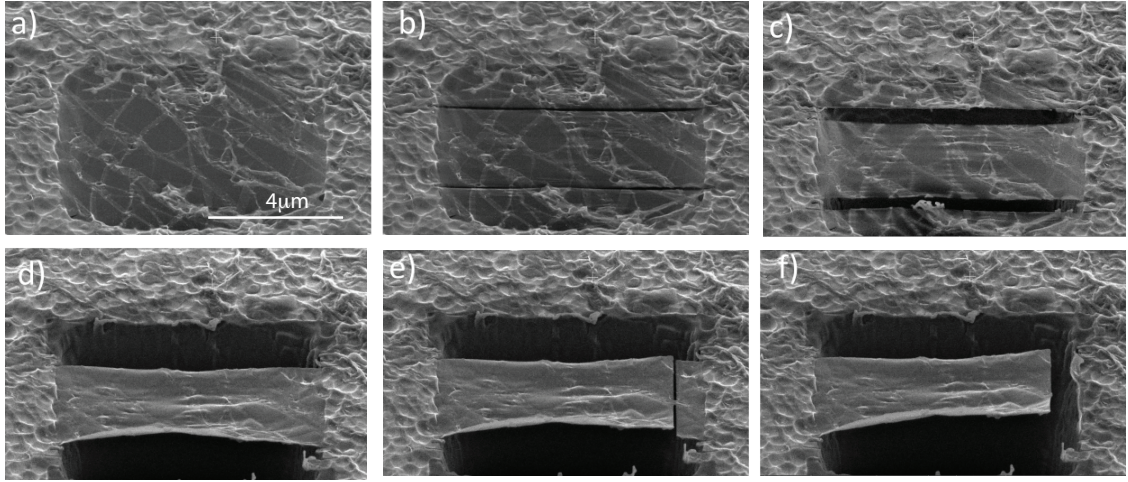

Figure S1: Sequential SEM snapshots illustrating the fabrication of graphene cantilevers. (a) Wrinkled graphene is primarily transferred over microcavities, the membrane is then sculpted to form a clamped-clamped beam in (b), (c), and (d), and eventually a cantilever as shown in (e) and (f).

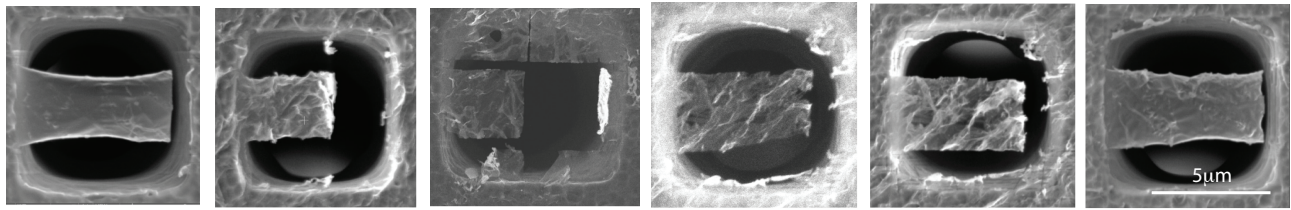

Figure S2: SEM images of a number of graphene cantilevers. The left image corresponds to the cantilever discussed in the main text.

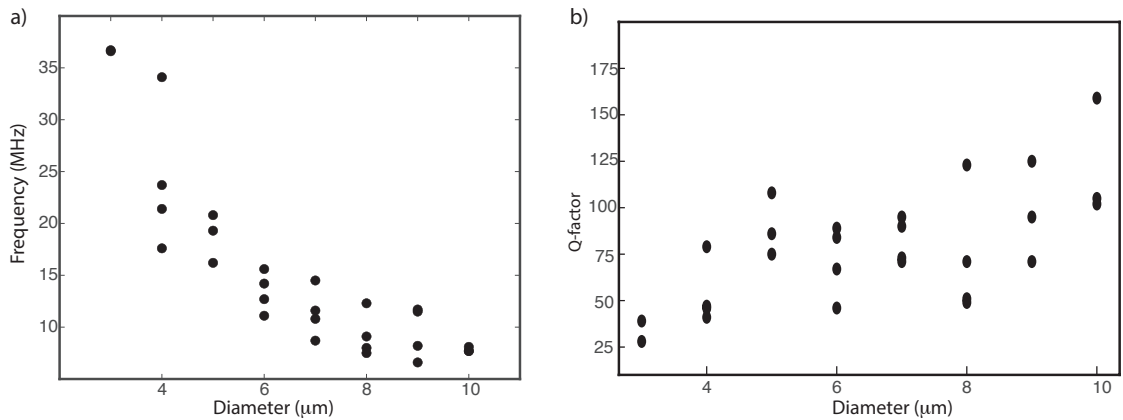

Figure S3: Diameter-dependent fundamental resonance (a) and Q-factor (b) of wrinkled graphene drums.

2

# S3. Modelling the Effects of Bending Stiffness

In this section we give details of the mathematical modelling performed to quantify the effect of bending stiffness in both the nano-indentation and resonance settings. Each of these uses models assuming that the graphene is clamped over a circular hole, simplifying the calculations by the assumption of axisymmetry (no azimuthal/hoop dependence). In the main text we compare experimental results for a rectangular hole of dimensions $w \times d$ with a circular hole of the same area, i.e. with effective radius

$$R_{\text{out}} = \left(\frac{wd}{\pi}\right)^{1/2}. \tag{S3.1}$$

In both cases, we shall model the graphene as an elastic sheet (of bending stiffness $\kappa$ and mass $\rho\bar{h}$ per unit area) that is clamped at $r = R_{\text{out}}$ and subject to a uniform pretension $\sigma_{\text{pre}}$ and a normal pressure, $p(r,t)$. The vertical displacements of the sheet, $z = \zeta(r,t)$, then satisfy

$$\rho\bar{h}\frac{\partial^2 \zeta}{\partial t^2} = p(r,t) + \sigma_{\text{pre}} \nabla^2 \zeta - \kappa \nabla^4 \zeta \tag{S3.2}$$

with clamped boundary conditions

$$\zeta(R_{\text{out}}, t) = \left.\frac{\partial \zeta}{\partial r}\right|_{r=R_{\text{out}}} = 0. \tag{S3.3}$$

## S3.1 Nano-indentation

Chandler & Vella [2] considered the equilibrium (i.e. time independent) version of (S3.2), first rendering the force and vertical displacement dimensionless by letting:

$$\mathcal{F} = \frac{F}{2\pi R_{\text{out}} \times \sigma_{\text{pre}}}, \quad \Delta = \frac{d}{R_{\text{out}}}.$$

Their result for the dimensionless indentation stiffness of a cylindrical indenter, radius $R_{\text{in}}$, is most easily expressed in terms of the compliance, as

$$k(\mathcal{B};\mathcal{R})^{-1} = \frac{\Delta}{\mathcal{F}} = \log\frac{1}{\mathcal{R}} + \frac{\hat{K}_1 \hat{I}_0^R + \hat{I}_1 \hat{K}_0^R + \mathcal{R}\left(\hat{K}_0 \hat{I}_1^R + \hat{I}_0 \hat{K}_1^R\right) - 2\mathcal{B}^{1/2}}{\hat{I}_1^R \hat{K}_1 - \hat{I}_1 \hat{K}_1^R} \frac{\mathcal{B}^{1/2}}{\mathcal{R}}, \tag{S3.4}$$

where

$$\mathcal{R} = \frac{R_{\text{in}}}{R_{\text{out}}},$$

$$\hat{I}_j = I_j\left(\mathcal{B}^{-1/2}\right), \quad \hat{I}_j^R = I_j\left(\mathcal{R}\mathcal{B}^{-1/2}\right), \tag{S3.5a}$$

$$\hat{K}_j = K_j\left(\mathcal{B}^{-1/2}\right), \quad \hat{K}_j^R = K_j\left(\mathcal{R}\mathcal{B}^{-1/2}\right) \tag{S3.5b}$$

and $I_j(\cdot)$ and $K_j(\cdot)$ are the modified Bessel functions of the first and second kinds (respectively) and order $j$.

A plot of the dimensionless stiffness $k(\mathcal{B};\mathcal{R})$ as given by (S3.4) is shown in fig. S4a for $\mathcal{R} = 2.2 \times 10^{-3}$ (based on $R_{\text{in}} = 10$ nm, $R_{\text{out}} = 8/\sqrt{\pi}$ μm $\approx 4.5$ μm). This plot demonstrates that (S3.4) recovers other asymptotic results in the appropriate limits. Of particular relevance in this paper are the cases of a perfect membrane [5] ($\mathcal{B} = 0$) and small (but finite) bending stiffness [2]:

$$k = \frac{\mathcal{F}}{\Delta} \approx \begin{cases} \left[\log(1/\mathcal{R})\right]^{-1}, & \mathcal{B} = 0 \\ \left(\gamma - \log 2 - \frac{1}{2}\log \mathcal{B}\right)^{-1}, & \mathcal{R}^2 \ll \mathcal{B} \ll 1, \end{cases} \tag{S3.6}$$

where $\gamma \approx 0.577$ is the Euler–Mascheroni constant. Note that the result for $\mathcal{R}^2 \ll \mathcal{B} \ll 1$ is valid only for intermediate bending stiffnesses, and does not recover the large $\mathcal{B}$ behaviour[1].

Comparing the expression in (S3.4) to experiments is difficult as we do not know the pretension $\sigma_{\text{pre}}$, which is used to convert the measured indentation stiffness, $K = F/\delta$, to a dimensionless stiffness, $k = \mathcal{F}/\Delta$. However, based on typical values for pristine graphene [3] (namely $\kappa \approx 1.2$ eV, $\sigma_{\text{pre}} \approx 0.1$ N/m) we would expect $\mathcal{B}_{\text{pristine}} \approx 10^{-7} \ll 1$. Since $\mathcal{B}_{\text{pristine}} \ll \mathcal{R}^2$, figure S4a suggests that the dimensionless stiffness is dominated by the indenter size, and hence is approximately independent of $\mathcal{B}$. Hence, a reasonable estimate of $\sigma_{\text{pre}}$ is

$$\sigma_{\text{pre}} \approx \frac{K_{\text{expt}} \log(1/\mathcal{R})}{2\pi}. \tag{S3.7}$$

---

[1] This intermediate asymptotic regime emerges because the effect of the zero slope boundary condition at the center persists over a horizontal length scale proportional to the bendocapillary length $\ell_{bc} = (\kappa/\sigma_{\text{pre}})^{1/2}$; this has the effect of introducing a 'virtual' indenter radius of typical size $R_{\text{eff}} \sim \ell_{bc}$ and hence indentation takes place with an effective dimensionless indenter size $\mathcal{R}_{\text{eff}} \sim \ell_{bc}/R_{\text{out}} = \mathcal{B}^{1/2}$. In this way, the $\log \mathcal{B}$ term in (S3.6) plays the same role as the $\log(1/\mathcal{R})$ term when $\mathcal{B} = 0$.

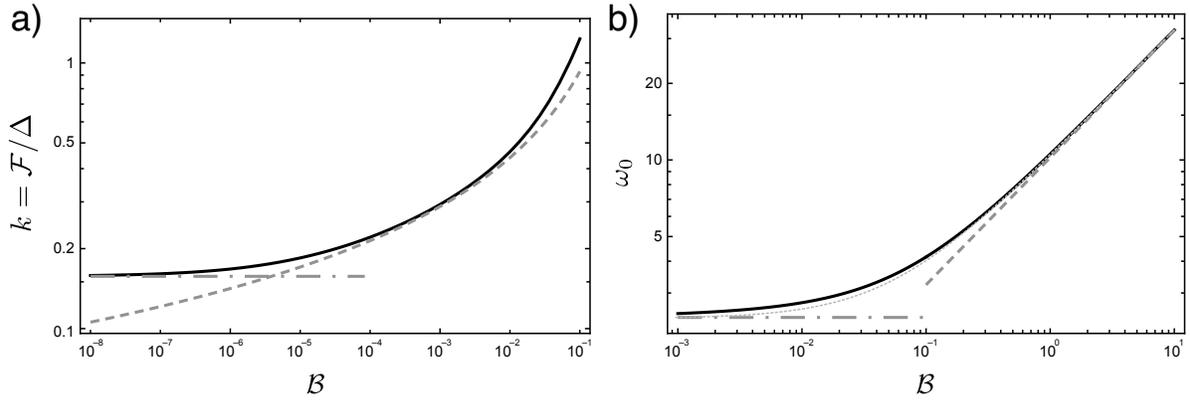

Figure S4: Plots of dimensionless predictions for the indentation stiffness and resonance frequency of plates with finite dimensionless bending stiffness $\mathcal{B}$. (a) The dimensionless indentation stiffness $k = \mathcal{F}/\Delta$ with dimensionless indenter size $\mathcal{R} = 2.2 \times 10^{-3}$ (solid curve). The asymptotic results from (S3.6) are shown for both $\mathcal{R}^2 \ll \mathcal{B} \ll 1$ (dashed curve) and $\mathcal{B} = 0$ (dash–dotted line) — note that the former only holds for intermediate values of $\mathcal{B}$, as expected. (b) The theoretical prediction accounting for both bending stiffness and pre-tension (solid curve) recovers the membrane limit (S3.12) (dash-dotted line) as $\mathcal{B} \to 0$; as $\mathcal{B} \to \infty$, the results recover the bending–dominated limit (S3.13) (dashed line). The approximate expression (S3.14) from ref. [1] is shown by the light gray dotted curve.

In our experiments, $\log(1/\mathcal{R}) \approx 6.11$ and so the proportionality constant $\log(1/\mathcal{R})/(2\pi) \approx 0.97$ — to a very good approximation $\sigma_{\mathrm{pre}} = K_{\mathrm{expt}}$ and so we assume that the effective tension $\sigma_{\mathrm{eff}} = K_{\mathrm{expt}}$. Our experiments give $K_{\mathrm{expt}} \approx 0.1$ N/m and hence

$$\sigma_{\mathrm{pre}} \approx 0.1 \text{ N/m}.$$

This is the value of the effective pretension, $\sigma_{\mathrm{pre}}$, that is used throughout the paper.

The experiments of the main paper further show a relative increase in $K_{\mathrm{expt}}$ of around a factor of 2 in the "surfactant-mediated" samples, as compared to the "surfactant-avoided" samples. This relative increase suggests that the dimensionless bending stiffness $\mathcal{B} \approx 1.6 \times 10^{-3}$, which, assuming that the pretension does not change between the two cases, would suggest that

$$\kappa_{\mathrm{surfactant}} \approx 1.6 \times 10^{-3} \times \sigma_{\mathrm{pre}} R_{\mathrm{out}}^2 \approx 2 \times 10^5 \text{ eV}.$$

## S3.2 Resonant frequency

To determine how the resonant frequency varies with $\mathcal{B}$, we focus on the limit in which $p(r,t) = 0$ (the sheet is free to vibrate) and incorporate the inertia of the sheet. As before, we non-dimensionalize lengths by $R_{\mathrm{out}}$, i.e. we let $\tilde{r} = r/R_{\mathrm{out}}$, and use the tension–inertia balance to determine the relevant time scale

$$t_* = \left(\frac{\rho \bar{h} R_{\mathrm{out}}^2}{\sigma_{\mathrm{pre}}}\right)^{1/2}.$$

Introducing a dimensionless time

$$\tilde{t} = t/t_*,$$

eqn (S3.2) then becomes:

$$\frac{\partial^2 \zeta}{\partial \tilde{t}^2} = \tilde{\nabla}^2 \zeta - \mathcal{B}\tilde{\nabla}^4 \zeta \tag{S3.8}$$

where $\tilde{\nabla}$ is the usual $\nabla$ operator, expressed in dimensionless variables.

We seek a dimensionless solution $\zeta(\tilde{r}, \tilde{t}) = F(\tilde{r})e^{i\omega \tilde{t}}$ and, suppressing $\tilde{\ }$, we find that $F(r)$ satisfies the eigenvalue problem

$$\omega^2 F = \mathcal{B}\nabla^4 F - \nabla^2 F \tag{S3.9}$$

with $F(1) = F'(1) = 0$ and finiteness conditions at the origin. We find that the general solution that remains finite as $r \to 0$ is

$$F(r) = AJ_0(\lambda_-^{1/2} r) + BI_0(\lambda_+^{1/2} r)$$

where $J_0(\cdot)$ is the Bessel function of the first kind and zeroth order, $I_0(\cdot)$ is the modified Bessel function of the first kind and zeroth order and

$$\lambda_\pm = \frac{1}{2\mathcal{B}}\left[\left(1 + 4\mathcal{B}\omega^2\right)^{1/2} \pm 1\right]. \tag{S3.10}$$

The boundary conditions can only be solved with non-trivial solution, $A, B \neq 0$, if a solvability condition

$$\lambda_+^{1/2} J_0(\lambda_-^{1/2}) I_1(\lambda_+^{1/2}) + \lambda_-^{1/2} I_0(\lambda_+^{1/2}) J_1(\lambda_-^{1/2}) = 0 \tag{S3.11}$$

4

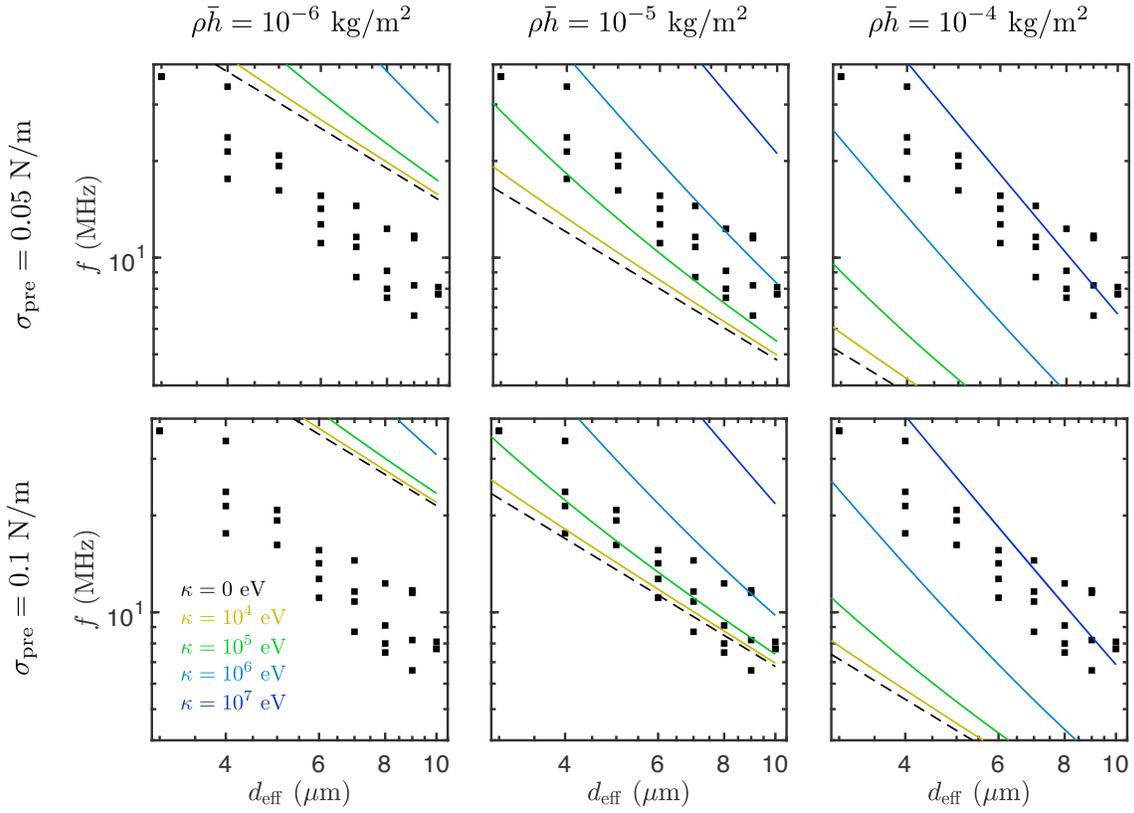

Figure S5: The dependence of the predicted dimensional resonance frequency on the effective hole diameter, $d_{\mathrm{eff}}$, assuming a range of the mass per unit area, $\rho\bar{h}$, and two values of the effective pre-tension $\sigma_{\mathrm{pre}} = 0.05\,\mathrm{N\,m^{-1}}$ or $\sigma_{\mathrm{pre}} = 0.1\,\mathrm{N\,m^{-1}}$. Theoretical predictions are shown as solid curves for a range of dimensional bending stiffnesses $\kappa$, as indicated in the legend. The dashed line shows the prediction for negligible bending stiffness, $\mathcal{B} = 0$, while experimental results are shown by the solid points.

is satisfied; this is the case only for certain values of the angular frequency $\omega$ as $\mathcal{B}$ varies. The smallest of these values of $\omega$, denoted $\omega_0(\mathcal{B})$, is the fundamental resonance frequency.

The function $\omega_0(\mathcal{B})$ must be determined numerically by solving (S3.11) with the $\lambda_\pm$ as defined in (S3.10). This numerical solution is shown in fig. S4b. It can be shown that in the limit $\mathcal{B} \ll 1$ the solution of (S3.11) is

$$\omega_0 \approx 2.405, \qquad (S3.12)$$

while for $\mathcal{B} \gg 1$

$$\omega_0 \sim 10.216\,\mathcal{B}^{1/2}. \qquad (S3.13)$$

These limits recover known results for the tension-dominated and bending-stiffness dominated [1] resonance of a sheet, respectively, which are also shown in fig. S4b. They are also close to the composite expression suggested by Castellanos-Gomez et al. [1], which in our notation reads:

$$\omega_0(\mathcal{B}) \approx \left[10.216^2\,\mathcal{B} + 2.405^2\right]^{1/2}. \qquad (S3.14)$$

This relationship is shown for completeness in fig. S4b.

Since the dimensionless angular frequency $\omega$ is related to the dimensional oscillation frequency $f$ by $f = \omega/(2\pi t_*)$, we can also use the theoretical results presented in fig. S4b to give predictions for the dimensional resonance frequency as a function of well size. To do this, we need estimates of the pretension $\sigma_{\mathrm{pre}}$ and areal mass density $\rho\bar{h}$ of the membrane (to determine the time scale $t_*$), as well as the bending stiffness $\kappa$. To see the effect of these two parameters, figure S5 shows results with a range of values of all three parameters. These results suggest that our experimental results are compatible with $\rho\bar{h} = 10^{-5}\,\mathrm{kg\,m^{-2}}$, $\sigma_{\mathrm{pre}} = 0.1\,\mathrm{N/m}$, and bending stiffness $\kappa = 2 \times 10^5$ eV; these are the values used in the corresponding figure in the main text (which also includes other values of $\kappa$ to highlight the effect of variation in $\kappa$ alone).

# Supplementary References



\end{document}